# The Disclosure of University Research for Third Parties:

# A Non-Market Perspective on an Italian University

Michelina Venditti,[a] Emanuela Reale,[b] & Loet Leydesdorff [c]

**Abstract**

Nations, universities, and regional governments commit resources to promote the dissemination of scientific and technical knowledge. One focuses on knowledge-based innovations and the economic function of the university in terms of technology transfer, intellectual property, university-industry-government relations, etc. Faculties other than engineering or applied sciences, however, may not be able to recognize opportunities in this "linear model" of technology transfer. We elaborate a non-market perspective on the third mission in terms of disclosure of the knowledge and areas of expertise available for disclosure to other audiences at a provincial university. The use of ICT can enhance communication between actors on the supply and demand sides. Using an idea originally developed in the context of the Dutch science shops, the university staff was questionnaired about keywords and areas of expertise with the specific purpose of disclosing this information to audiences other than academic colleagues. The results were brought online in a thesaurus-like structure that enables users to access the university at the level of individual email address. This model stimulates variation on both the supply and demand side of the innovation process, and strengthens the accessibility and embeddedness of the knowledge base in a regional economy.

---

[a] Gabriele d'Annunzio University, Chieti-Pescara, Department of Business Administration and Management, Italy; mvenditti@unich.it
[b] CERIS CNR, Via dei Taurini, 19, 00185 Rome, Italy; e.reale@ceris.cnr.it
[c] University of Amsterdam, Amsterdam School of Communication Research (ASCoR), Kloveniersburgwal 48, 1012 CX Amsterdam, The Netherlands; loet@leydesdorff.net

## 1. Introduction

In 1998, the UK government introduced wealth creation as a “third mission” for universities in addition to teaching and research. The initiative was incentivized with £50 million annually (Klein, 2002; Martin & Tang, 2007; Molas-Gallart *et al.*, 2002). The idea of organizing a Third Mission, however, is much older; for example, the land-grant universities in the United States were organized in the 19th century (Clark, 1998; Etzkowitz, 2002). Changing an institution’s mission may lead to debates among academics, governments, and stakeholders since the transformation of universities as organizations requires the development of a set of functions linked to the new mission.

In recent decades, the emphasis on knowledge-based innovations has featured the economic function of the university as focal among the third missions. The “Third Mission,” for example, was mainly implemented in terms of technology transfer offices, intellectual property, “valorization” programs, university-industry-government relations, etc. (e.g., Sanchez & Elena, 2006; Rothermel, 2007). For example, the Bayh-Dole Act in the United States called for universities to transfer knowledge to business and stimulated the further development of “the entrepreneurial university” through the protection of intellectual property. Adopting this discourse, policy makers at various levels tend to assume the linear model of innovation. In our opinion, the linear model (of technology push) limits the possibilities for developing university-industry and university-third party relations in terms of processes of mutual learning and possible adjustments.

More recently, science policies have also developed a focus on knowledge *dissemination.* In 2007, for example, the United States introduced the Scientific Communications Act (law HR1453), which authorized the National Science Foundation to spend $10M annually to the education of communication of scientists annually during the period 2008-2012. Other countries, including emerging ones such as India and China, have also committed considerable resources to promote the dissemination of knowledge and the relationship between science and society (Greco, 2007). In terms of models of science communication, however, this approach assumes a deficit model: the knowledge would already be available, and has only to be communicated more efficiently to the end-user. This approach, in our opinion, also prevents the learning process from occurring on both sides (e.g., Wynne, 1991; 1995).

Dissemination of knowledge was also central to the Berlin Declaration on Open Access to Knowledge in the Sciences and Humanities, which state that "disseminating knowledge is only half complete if the information is not made widely and readily available to society. New possibilities of knowledge dissemination not only through the classical form but also and increasingly through the Open Access paradigm via the Internet have to be supported." Following this approach, many universities have now set up institutional repositories in which the scientific outputs of professors, researchers, fellows, graduate and doctoral students, and technically graduated staff are stored. However, the main purpose is to enable universities to improve their visibility and reputation with external audiences, by highlighting the bulk and diversity of results produced (Reale *et al*., 2011). Thus, one "sends" information, but one does not "hear" the questions.

Most case-study research about university-industry-government relations and technology transfer focuses on "best practices" such as MIT and Stanford or some European entrepreneurial universities (e.g., Etkzowitz *et al.*, 2000; Jacob *et al.*, 2003; Saxenian, 1999). In our opinion, one should not narrow the "third mission" *beforehand* to best-practices and industrial demand for innovation, but pay attention to efforts for improving the practices of communication between academia, industry, and third parties. In a knowledge-based economy, variation is just as important on the demand side as it is on the supply side (Laredo, 2007). Before one focuses on success stories about "building bridges," the rich varieties on both banks of the river can be made visible for different audiences so that more options for innovations can be explored.

Perhaps, it is useful to keep in mind that only approximately 10% of the innovative ideas in advanced industries lead eventually to successful innovations. Leaving the pre-selection in this process exclusively to private demand thus seems counterproductive. The notion of "government" in university-industry-government relations provides room for additional democratization in terms of access to research capacities. Interaction processes between supply and demand provide room for a wider participation by citizens and their organizations in a non-linear process of education, transfer and exchange, and utilization (Mowery & Rosenberg, 1979).

In this study, we further develop the interface between supply and demand by elaborating on an idea that originated in the context of the Dutch science shops in the late 1980s. One of us participated in a project at the time called the development of a “Science Bank” (Leydesdorff, 1988). In collaboration with the Boards of the two Amsterdam universities all tenured staff were questionnaired when gathering the yearly information for the Annual Reports with two additional questions:

1. Can you provide specific keywords for disclosing your current research to third parties?
2. Do you have other expertise (from previous research projects) which can be made relevant for third parties?

In a pilot study, it had first been found that the order of these questions is sensitive: asking the latter question first tends to obscure the answering to the first one additionally. At the time (1988), 4,495 keywords and 810 expertise specifications were collected from 894 research projects. The results and a search engine were made available on a CD-Rom. Users could query the system and then receive relevant telephone numbers and address information of staff members. The results were tested by and received a positive evaluation from the Innovation Center of the regional Chamber of Commerce. However, the project was subsequently terminated due to lack of funding.

We further elaborate this cognitive perspective on the third mission. In contrast with institutional relations, bridges in the cognitive dimension can be constructed between social partners even before one interacts socially; this quality makes it possible to create variation in the relationships, which can yield fruitful ideas. By further developing the relevant repertoires, one can facilitate the translations needed by articulating common interests despite differences in the semantics. An ICT tool can help in this process of communication by providing intelligent support to the translations. By providing this data, furthermore, the third mission of the university can be made an empirical subject of study since data is made available.

## 2. Theoretical framework

The concept of a Third Mission developed from the pressure for change that universities have been experiencing since the 1980s. In the context of reindustrialization policies and the emerging knowledge-based economy, agencies such as the OECD have argued for a more engaged role of academia in the economy and society (Freeman, 1982; Rothwell & Zegveld, 1981). The need to improve the university's performance towards its societal and regional environments has in the meantime widely been acknowledged.

During the nineties, the notions of national innovation systems (Lundvall, 1992; Nelson, 1993), modes of knowledge production (Gibbons *et al.*, 1994), and the triple helix model (Eztkowitz & Leydesdorff, 1997; 2000) contributed to a shift toward considering universities as key actors within the knowledge production system (e.g., Godin & Gingras, 2000). The integration into the local environments can be considered analytically different from integration into the global environment of international publishing and patenting. Accordingly, the communication mechanisms are very different at local and global levels.

The Third Mission is meant to change the way universities are embedded into regions and communities. In the so-called Russell Report (Molas-Gallart *et al.*, 2002), the Third Mission was first defined in terms of a set of third-stream activities, which rest on the capabilities—what universities *have* in terms of knowledge capabilities and facilities—and on the activities—what universities *do* in terms of research, teaching, and communication. These activities can then be described using a set of indicators “representing the main ways through which universities engage potential non-academic users and beneficiaries”.

Spaapen *et al.* (2007) elaborated this indicator approach into “evaluation of research in contexts.” From this perspective of evaluation, one focuses on reporting the interactions, that is, communication and collaboration, between researchers and their societal audiences. Different ways of interactions can then be distinguished such as common agenda-setting, collaborative research, communication and dissemination of research outcomes, the use of knowledge for productive purposes, etc. In our opinion, these indicators can be considered as “proxies” for understanding quality and relevance of research for science and society because they indicate opportunities for the co-construction of knowledge between different academic and non-academic actors. The focus in these indicator studies, however, has remained on reporting “Third Mission” activities for evaluation in the hope that this articulation would garner more attention for such processes in research practices.

The further development of Third-Mission indicators has also been the subject of a number of European Projects such as E3M (“European Indicators and Ranking Methodology for University Third Mission")[1] and SIAMPI ("Social Impact Assessment Methods for research and funding instruments through the study of Productive Interactions"),[2] or the methodology developed within the OEU (Observatory of the European Universities)[3] that identify four societal dimensions and four economic dimensions for shaping indicators of third-mission activities.

Given the context of evaluation in indicator research, these various attempts to deal with the third mission tend to focus on specific agency (first and foremost firms) and specific types of activities (patenting, technology transfer, mobility, contracts with industry, advice to government, public understanding, participation in social life). This implies that many activities and results, especially in "soft" disciplines such as the humanities and the social sciences, have been mostly neglected, because they have less chance of making an "impact"; they have a reduced capacity to produce "valid" knowledge that can be "valorized" on a market.

A different perspective was proposed by Laredo (2007). He suggested shifting the discourse from three institutional missions of universities to three functions, namely (*i*) the tertiary education of large numbers of students, (*ii*) professionally specialized higher-education and research, and (*iii*) academic training and research output. These three functions correspond to different arenas: the first function corresponds to the local arena , the second to the arena of the various professions , and the third to the global level of the development of the sciences. Each university performs as a mix of these three functions and profiles of universities are constructed according to contingent and historical contexts. Third-mission activities can typically be linked to the positioning of universities between these three functions, and new perspectives can be found by combining the different disciplinary fields with these functions. A non-market perspective of the third mission implies the broadening of the information disclosed, and the need to further improve the number of channels for the exchange. From this perspective, the extent to which ICT can provide added value for enhancing this communication process is an important item to be further explored.

[1] www.e3mproject.eu/
[2] www.siampi.eu/Pages/SIA/12/625.bGFuZz1FTkc.html

## 3. Methods and data

### *3.1 The Gabriele d'Annunzio University of Chieti-Pescara*

The Gabriele d'Annunzio University of Chieti-Pescara will provide our domain. The university was founded in 1965 as a "free" university by a consortium of three municipalities and their respective provinces in the Abruzzi region of Italy. This Free University was turned into the Public State University "Gabriele d'Annunzio" (UdA) in 1982 with premises in Chieti (Faculty of Medicine, and Faculty of Philosophy and Literature), Teramo (Faculty of Law and Faculty of Political Science), Pescara (Faculty of Economics and Trade, Faculty of Foreign Languages, Faculty of Architecture), and the Rectory in Chieti. In 1993, however, Teramo decided to leave UdA and found its own university.

UdA can be considered a medium-sized university among the 95 Italian universities: it has 12 faculties with 27,092 students, a body of 728 professors and researchers, structured in 23 departments. Table 1 shows the distribution of the academics and students across the 12 faculties of UdA.

[3] www.prime-noe.org

**Table 1** Frequency table of UdA academics and UdA students among the 12 Faculties

| *UdA Faculties* | *Number of academics* | *Relative frequency of academics (%)* | *Number of students* | *Relative frequency of students (%)* |
|---|---|---|---|---|
| Medicine | 204 | 28.0 | 3,319 | 12.3 |
| Architecture | 79 | 10.9 | 2,755 | 10.2 |
| Economics | 78 | 10.7 | 3,211 | 11.9 |
| Humanities | 55 | 7.6 | 1,558 | 5.8 |
| Linguistics | 55 | 7.6 | 2,317 | 8.6 |
| Pharmacy | 52 | 7.1 | 3,448 | 12.7 |
| Management Science | 49 | 6.7 | 2,210 | 8.2 |
| Psychology | 38 | 5.2 | 3,849 | 14.2 |
| Social Science | 35 | 4.8 | 1,027 | 3.8 |
| Sport | 28 | 3.8 | 2,030 | 7.5 |
| Mathematics, Physics, Natural Science | 28 | 3.8 | 324 | 1.2 |
| Education Science | 27 | 3.7 | 1,044 | 3.9 |
| ***Sum*** | *728* | 100% | *27,092* | 100% |

In 2011, the university entered a process of reorganizing its internal structure and bodies to comply with the provisions stated in the so-called Gelmini Reform Act (Law 240/2010) that changes the system of governance, institutional arrangements, and administrative management of Italian universities. As a consequence UdA will have in the future departments with a minimum size of 35 professors and tenured staff. The “new” departments are the primary centres of scientific research and teaching.

Furthermore, UdA signed the above-mentioned Berlin Declaration on Open Access to Knowledge in the Sciences and Humanities in 2005. In this context, UdA stimulates the academic staff to submit degree theses, PhD theses, scientific publications, preprints, and didactic materials for students to four repositories. In the new statute one foresees (in article 12) that UdA will adopt the principles of full and open access to scientific literature and promote free dissemination of the results on the web in order to ensure the widest possible distribution.

On the homepage of the university (www.unich.it), a banner was furthermore inserted to highlight UdA's own scientific journal for reporting its activities in the different scientific areas. "News and events" links provide videos about significant research activities, participation in congresses, education, and internationalization activities. However, the main audiences of these activities have remained academic colleagues and students within academia.

### *3.2 Methodology*

The project is based on a collaboration agreement between the Department of Management and Business Administration of UdA and the Union of the Chambers of Commerce in the region. The Union of the Chambers of Commerce is a non-profit institution with categorical associations such as Confindustria (Union of the representatives of industries), Confapi (Union of representatives of small business organizations), CNA (Confederation of the crafts) as its members. A formal contract and letter of intent was exchanged between the Council and the Department of Management and Business Administration in July 2011.

In the period May 2011-April 2012, a web-based survey was run addressing all the academic staff of the UdA. The objective was to construct the science knowledge bank, through the disclosure of keywords related to the research of individual academics, and disclosure of their expertise. The survey was sent online, though the use of the freeware facility "Survey Monkey" (at www.surveymonkey.com), to all the 728 academics of UdA. In anticipation to the survey we sent a letter to the Rector's office, the General Manager, the Deans of the Faculties and Heads of the Departments, with the aim of introducing the project and asking for their collaboration in supporting the participation and involvement of academic staff in the survey. Furthermore, the survey was preceded by a pre-test: copies were handed out to 16 UdA academics, to verify that the questions were worded clearly.  After this pre-test, we agreed on the following five questions:

1. Name, surname, and academic position in the university;
2. Departmental address;
3. Scientific area (in Italian the so-called: *settore scientifico-disciplinare,* or S.S.D.);
4. Can you provide specific keywords for the disclosure of your current research to third parties in society?
5. Do you perhaps have other expertise (from previous research projects) which can be made relevant for third parties?

The cover letter contained a recommendation of the Head of the Department of Management and Business Administration, Professor Augusta Consorti. The population ($N$ = 728) was defined on the basis of the website of the Italian Ministry of University and Scientific

Research (www.miur.it) and classified into full professors (in Italian the so-called *ordinari confermati e non*), associated professors (*associati confermati e non*), senior lecturers (*ricercatori confermati*), lecturers (*ricercatori non confermati*), and assistant professors (*ricercatori a tempo determinato*).

The analysis of the answers to questions 4 and 5, provided by the participating academics, enables us to collect keywords about the research currently underway, and potential areas of expertise from the past, in terms of the academics of UdA (question 1). The answers to questions 2 and 3 provide a basis for the statistical information.

### *3.3 Composition of the reference population*

The population of potential respondents is mostly composed of full professors (31.3%) followed by associate professors (28%), senior lecturers (27.5%), while the lecturers (9.5%), and the assistant professors (3.7%) represent limited shares of the population. The latter two categories have no tenure, but are on tenure-track. Using the 14 research areas prescribed in Italy by law (Ministerial Degree of October 4, 2000), the 728 academics of UdA are distributed as shown in Figure 1.

**Figure 1.** *Relative frequency table of UdA academic position per research area*

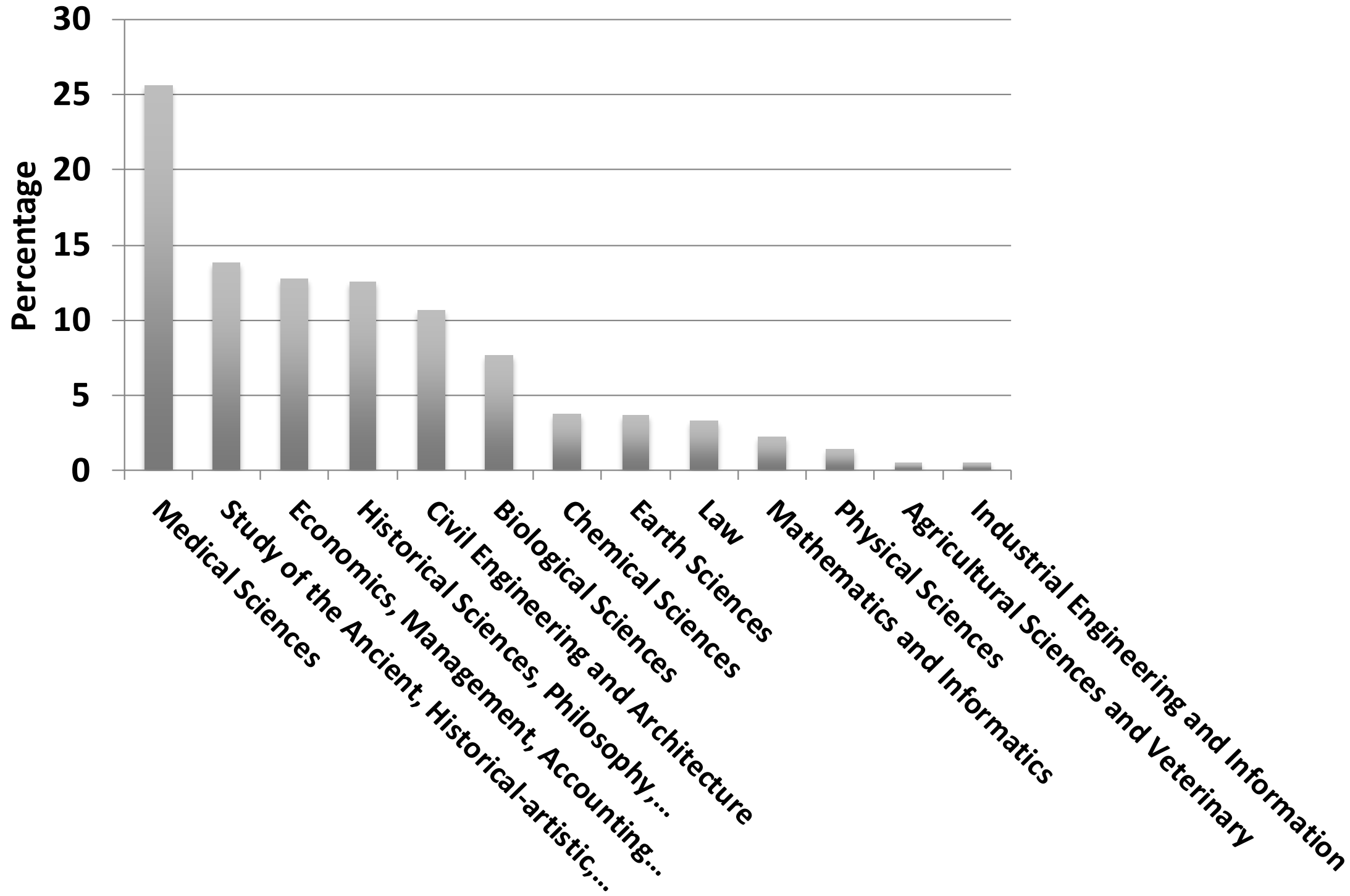

Figure 1 shows a prevalence of academics in the Medical Sciences (25.6%), followed by the Study of the Ancient, Historical-artistic, and Eastern Philology and Literature (13.8%), Economics, Management, Accounting and Statistics (12.7%), Historical Sciences, Philosophical, Education, Psychological, (12.5%), Civil Engineering and Architecture (10.6%). Given this profile—perhaps with the exception of the medical faculty and the department of civil engineering—this university cannot be expected to contribute strongly to traditional technology transfer.

## 4. Results

### *4.1 Responses*

After three rounds of email reminders, the total response was 220, that is, a response rate of approximately 30%. The responses are more or less equally distributed when compared with the reference population in terms of the categories (Table 2).

**Table 2.** *Table of frequency and percentage of response rates of UdA academics per position*

| *UdA Academic position* | *Frequency* | *Percentage* | *Percentage of respondents per position with respect to the reference population* |
|---|---|---|---|
| Full Professor | 61 | 29.1 | 26.7 |
| Associate Professor | 64 | 27.7 | 31.4 |
| Senior Lecturer | 61 | 27.7 | 30.5 |
| Lecturer | 25 | 11.4 | 36.2 |
| Assistant Professor | 9 | 4.1 | 33.3 |
| ***Sum*** | 220 | 100.0 | |

Table 3 provides the distribution of the respondents over the different research areas. Academics belonging to the areas of Social and Political Sciences (85.7%) and Economics, Management, Accounting and Statistics (63%) prevail, while the areas of Industrial Engineering and Information as well as the Agricultural and Veterinary Sciences are virtually absent among the respondents. The areas of Civil Engineering and Architecture (12.9%) and Medical Sciences (16%) are the next least represented fields among the respondents.

**Table 3**. *Frequency table and response percentage of UdA survey respondents per research area*

| *Academic respondents per research area* | *Frequency* | *Percentage* | *Percentage of respondents per research area with respect to the re-ference population* |
|---|---|---|---|
| 01: Mathematics and Informatics | 3 | 1.4 | 18.7 |
| 02: Physical Sciences | 3 | 1.4 | 30.0 |
| 03: Chemical Sciences | 10 | 4.5 | 37.0 |
| 04: Earth Sciences | 11 | 5.0 | 42.3 |
| 05: Biological Sciences | 12 | 5.4 | 21.8 |
| 06: Medical Sciences | 30 | 13.6 | 16.0 |
| 07: Agricultural Sciences and veterinary | 0 | 0.0 | 0.0 |
| 08: Civil Engineering and Architecture | 10 | 4.5 | 12.9 |
| 09: Industrial Engineering and Information | 1 | 0.4 | 100.0 |
| 10: Study of the Ancient, Philological and Literary, Historical-artistic and Eastern | 23 | 10.4 | 23.0 |
| 11: Historical Sciences, Philosophical, Education, Psychological, Demo-anthropological, Geography, Sports | 31 | 14.2 | 34.1 |
| 12: Law | 10 | 4.5 | 41.7 |
| 13: Economics, Management, Accounting and Statistics | 58 | 26.5 | 63.0 |
| 14: Social and Political Sciences | 18 | 8.2 | 85.7 |
| ***Sum*** | 220 | 100.0 | |

In summary, one-third of the academic staff was responsive to the initiative and the response was proportionally distributed among the same age group, but unequally among the disciplines.

### *4.2 Keywords and expertise*

The 220 respondents communicated 1,115 keywords and 523 areas of expertise. The keywords were provided from the minimum of just a single keyword to a maximum of 32 keywords per respondent. Respondents also provided areas of expertise, ranging from one single expertise to a maximum of 14 areas of expertise. All these distributions are skewed (as can be expected; cf. Ijiri and Simon, 1977; Leydesdorff, 1989).

The distribution of the keywords per research area is dominated by respondents in Economics and Statistics, due to the fact that most respondents belonged to this discipline.[4] A large majority of the respondents (91%) gave three keywords. In the area of the humanities, 330 keywords were provided by respondents belonging to this area, with 46% of the respondents providing five keywords and 13% offering even seven keywords for the disclosure of her work. This response shows a wide array of activities relevant for third parties and an openness on the part of academics to communicating these resources with the surrounding social environment.

Seventy percent of the respondents were also able to indicate relevant areas of expertise. Respondents belonging to the areas of Economics and Statistics prevailed, but once again, academics from the humanities and the social sciences mentioned a large variety of relevant expertise in areas that cannot be considered relevant to technology transfer, but which may be very valuable for giving answers, insights, and perhaps solving problems in relation to social demands. As to the distribution over academic positions, most of the keywords were volunteered by junior scholars, while most of the areas of expertise were listed by the senior staff.

### *4.3 Analysis of the different keywords and areas of expertise*

After the data cleaning (avoiding repetitions, and unification of keywords in terms of spelling), the 220 responses provided us with 988 keywords and 494 areas of expertise. Most

keywords and areas of expertise are composed of phrases of two or more words, such as: “public transport,” “social capital,” “intellectual capital,” “*sociologia di Marx*”. They are also a mixture of English and Italian. Given the intention to disclose the research to wider audiences by following the wordings proposed for this purpose by the researchers themselves, we did not translate the keywords either from or to English or Italian, but left them in their original form. Certain researchers provided some keywords in English and others in Italian.

For the purpose of disclosure it was necessary to avoid false positives, that is, to remove words generated by the ICT system that had no clear significance as keywords. Further data cleaning and permutation of composed keywords led to 1,706 single-word keywords and 1,500 areas of expertise. Both these single words—which may overlap among researchers with different areas of expertise or who provided different keywords as phrases—and the original keywords and areas of expertise are included in a clickable index which was brought online at http://www.leydesdorff.net/pescara-chieti/index.htm (Figure 2).

[4] One respondent provided no keywords, but only areas of expertise.

**Figure 2**: *Index of keywords and expertise at the G. d'Annunzio University, Chieti-Pescara, 2011; at http://www.leydesdorff.net/pescara-chieti/index.htm.*

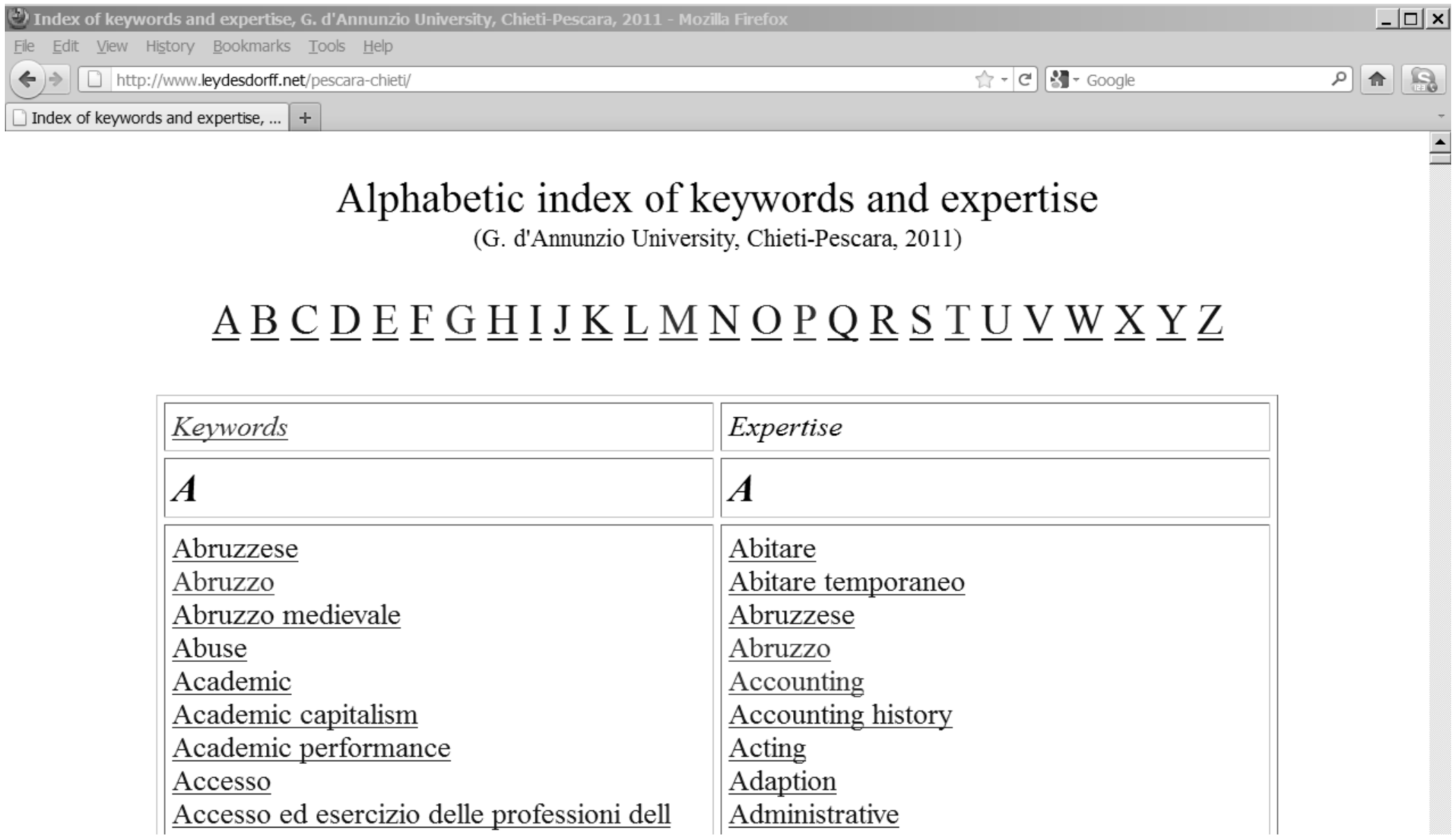

By clicking on any of the words, the user obtains access to a screen with the relevant researchers, their email addresses, other areas of expertise, and mentioned keywords in tabular form. Figure 3 shows this in greater detail, offering as an example the results for "accounting" as an area of expertise.

**Figure 3**: *Listing provided for "accounting" as an area of expertise (at http://www.leydesdorff.net/pescara-chieti/expert/accounting.htm).*

| *Name* | *Keywords* | *Expertise* |
|---|---|---|
| Daniela DI BERARDINO<br>< daniela.diberardino at unich.it > | - intangible assets<br>- creazione del valore<br>- corporate governance<br>- comunicazione economico-finanziaria | - accounting<br>- valutazione d azienda<br>- sviluppo progetti<br>- pianificazione<br>- controllo |
| Stefania MIGLIORI<br>< s.migliori at unich.it > | - crisis management<br>- corporate governance<br>- small and medium enterprises<br>- spin-off | - international accounting standards<br>- accounting<br>- accounting history |
| Francesco DE LUCA<br>< fdeluca at unich.it > | - principi contabili<br>- ias_ifrs<br>- standardizzazione contabile<br>- accounting<br>- controllo di gestione | - strategie aziendali<br>- corporate governance<br>- storia bilancio dello stato<br>- accounting<br>- business plan |

Figure 3 shows that, for example, one of the staff members thus retrieved has specific expertise concerning small and medium enterprises and spin-off processes. This combination of keywords enables users to make more informed selections, and address researchers with considerable precision. We expect this to be beneficial to both sides of this process of mediation.

We also experimented with a network visualization program for drawing a semantic map using these words. However, most formulations of the keywords and areas of expertise occur only once; the database reveals a wide variety of terms and expertise more than a (selective) structure. Among the 988 keywords, only 41 (4.1%) occur more than once; among the 494 areas of expertise, only 17 areas are repeated (3.4%). In other words, the keywords and areas of expertise mentioned show mainly variation, but not as a networked structure.

**Figure 4**: *Network visualization using keywords occurring more than once and department names (Fruchterman & Reingold, 1991; installation in Pajek).*

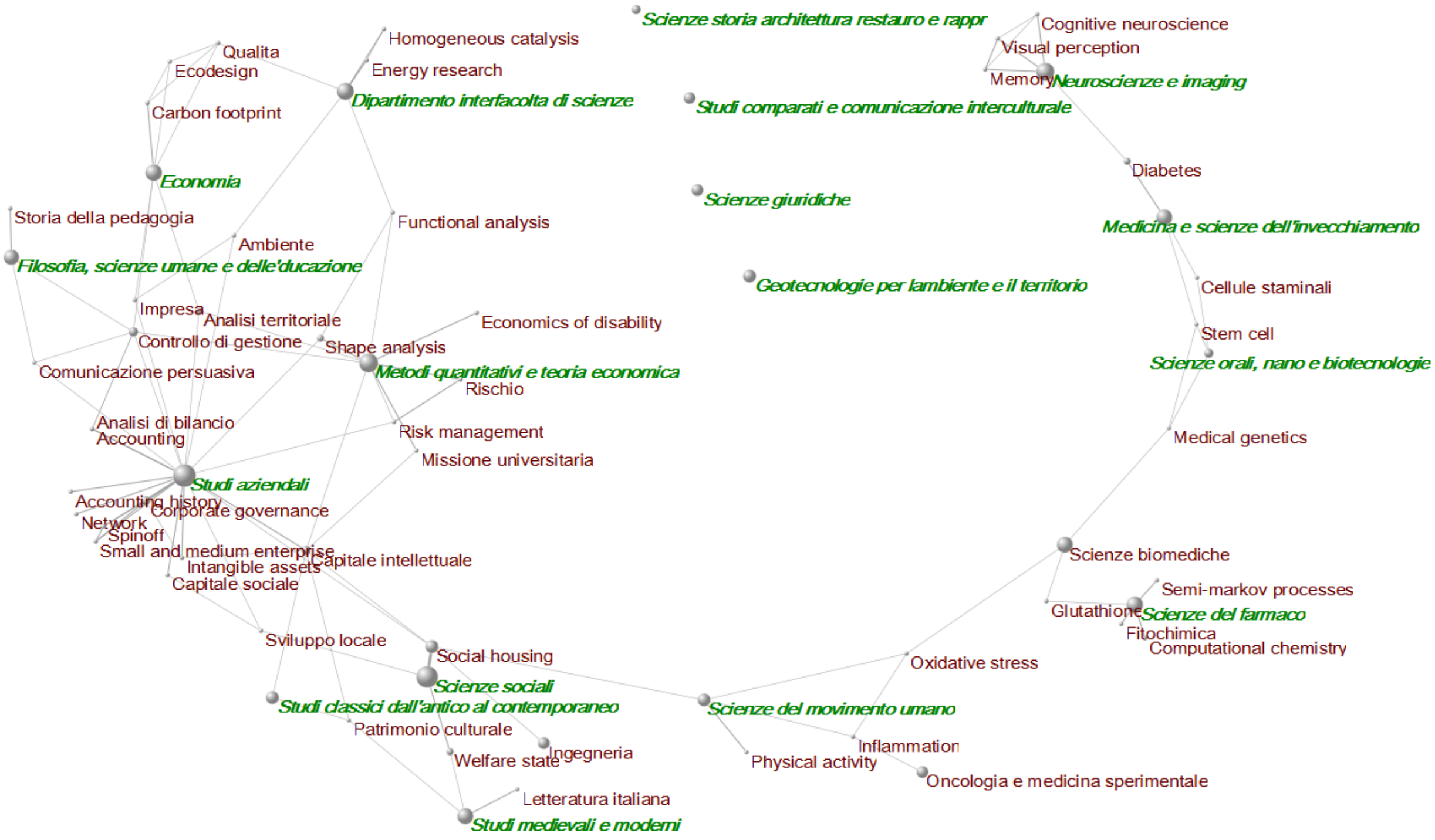

Figure 4 shows the structure based on the 41 keywords mentioned twice combined with the departmental names. Although such a structure can be made clickable (and partially is made clickable at http://www.leydesdorff.net/pescara-chieti/keywords.htm), the semantic content is not convincing: several of the departments remain unconnected in this representation. We therefore decided not to pursue the map-based disclosure and chose to use the index lists (which show the variation) instead. The alternative of using single occurrences in a mapping would overload the image, making it hard to read.

## 5. Conclusions and discussion

In this paper we have presented the results of a project proposing that the third mission of universities be made an empirical subject of study and not only one of legitimation in evaluations or an issue for normative debate. The results of this experiment in the Abruzzi Region (Italy), involving the Gabriele d'Annunzio University and the Union of the Chambers of Commerce of the region (Chieti, Pescara, Teramo, and L'Aquila) were made available for online users. In particular we reported on the development of a phase intended to build up and communicate the supply side of research and expertise of UdA's academics using ICT tools able to improve the way in which each academic can show his/her research topics and results to a broader audience.

The results show that an untapped reservoir of socially relevant material and expertise can be made visible by these relatively simple and cheap instruments. As we noted, the willingness to open up to the social environment seems negatively correlated to the traditional methods that have proven unsuitable for transferring and valorizing the knowledge of scientific specialties. The humanities, for example, score high on providing keywords, whereas the social sciences indicate many areas of expertise relevant for the surroundings. The participation from the side of the medical sciences was also considerable.

In our opinion, the generation and further development of a knowledge-based economy requires many more options for innovation than can be demanded for by private companies. Citizens can be more involved by raising public demand for innovation and the universities can provide the innovation system with the supply side information that is locally available. In the future, we envisage that the two lists (for supply and demand side, respectively, once the latter will be developed) will enable us to build a piece of artificial intelligence to guide users from the one domain to the other. The user would be able to access the system with keywords and find direct access to individual staff members who are interested in these issues. Thus defined, the "third mission" is no longer confined to what is happening in terms of "best practices," but in terms of what could be possible in terms of fruitful communication.